\documentclass{article}
\usepackage{spconf,amsmath,graphicx}

\usepackage{hyperref}
\usepackage{algorithm}
\usepackage{algpseudocode}
\usepackage{amsfonts}
\usepackage{longtable}
\usepackage{multirow}
\usepackage{xcolor}
\usepackage{adjustbox}
\usepackage{xspace}
\usepackage{booktabs}

\makeatletter
\newcommand\subpara[1]{%
\textbf{#1: }%
\@ifnextchar\par{\@gobble}{}}
\makeatother

\makeatletter
\let\oldlabel\label
\renewcommand\label[1]{%
\oldlabel{#1}%
\@ifnextchar\par{\@gobble}{}}
\makeatother

\makeatletter
\DeclareRobustCommand\onedot{\futurelet\@let@token\@onedot}
\def\@onedot{\ifx\@let@token.\else.\null\fi\xspace}

\title{
    Search Wide, Focus Deep: Automated Fetal Brain Extraction \\
    with Sparse Training Data
}

\name{
    Javid Dadashkarimi$^{1,2}$,   
    Valeria Pena Trujillo$^{1,3}$,
    Camilo Jaimes$^{1,3}$, 
    Lilla Zöllei$^{1,2}$, 
    Malte Hoffmann$^{1,2}$
}

\address{
    $^1$ Department of Radiology, Harvard Medical School, Boston, MA, USA \\
    $^2$ Martinos Center for Biomedical Imaging, Massachusetts General Hospital, Charlestown, MA, USA \\
    $^3$ Pediatric Imaging Research Center, Massachusetts General Hospital, Boston, MA, USA
}

\begin{document}

\maketitle

\begin{abstract}

Automated fetal brain extraction from full-uterus MRI is a challenging task due to variable head sizes, orientations, complex anatomy, and prevalent artifacts. While deep-learning (DL) models trained on synthetic images have been successful in adult brain extraction, adapting these networks for fetal MRI is difficult due to the sparsity of labeled data, leading to increased false-positive predictions. To address this challenge, we propose a test-time strategy that reduces false positives in networks trained on sparse, synthetic labels. The approach uses a breadth-fine search (BFS) to identify a subvolume likely to contain the fetal brain, followed by a deep-focused sliding window (DFS) search to refine the extraction, pooling predictions to minimize false positives.
We train models at different window sizes, synthesizing images from a small number of fetal brain label maps augmented with random geometric shapes. Each model samples diverse head positions and scales, including cases with partial or absent brain tissue. Our framework matches state-of-the-art brain extraction methods on clinical HASTE scans of third-trimester fetuses, and exceeds them by up to 5\% in terms of Dice in the second trimester as well as for EPI scans across both trimesters. Our results demonstrate the utility of the sliding-window approach in DL and of combining predictions from several models trained on synthetic images, progressively improving brain-extraction accuracy.
\end{abstract}

\begin{keywords}
fetal brain extraction; deep learning; CNN; MRI; synthetic data; sparse annotations
\end{keywords}

\section{Introduction}

Automated fetal-brain extraction from full-uterus magnetic resonance imaging (MRI) data is a critical task in medical image processing, often used in developmental research studies and for the detection of abnormalities~\cite{scheinost2023machine}. 
Recent approaches have increasingly used deep learning (DL), predominantly leveraging convolutional neural networks (CNNs)~\cite{salehi2018real,ebner2020automated,faghihpirayesh2024fetal}, to address the challenges of fetal-MRI analysis: variable head sizes and orientations within a large field of view (FOV), distracting fetal and maternal anatomy, and a high prevalence of artifacts, such as incoherent geometry caused by subject motion between the acquisition of individual slices. These challenges frequently lead to inaccurate predictions~\cite{faghihpirayesh2024fetal,tajbakhsh2020,ciceri2023review}.

A recent class of DL methods synthesizes extremely diverse images from anatomical label maps to alleviate the need for manually labeled training data~\cite{hoopes2022synthstrip,shang2022learning,billot2023synthseg,zalevskyi2024improving}.
In addition, the high variability of the generated data helps networks generalize across populations and MRI contrasts unseen at training~\cite{valabregue2024comprehensive}.
Unfortunately, the lack of publicly available label maps that cover the fetal brain, body, \textit{and} maternal anatomy limits the applicability of this strategy to fetal MRI~\cite{heaven2019deep, jacob2024dcsm}.
While synthetic random-shape labels have proven beneficial for segmentation~\cite{dey2024anystar} and registration~\cite{hoffmann2021synthmorph}, completing sparse anatomical labels with geometric shapes to learn fetal-brain extraction remains under-explored.

In this paper, we build on prior work that uses the synthesis strategy to learn skull-stripping of adult-brain MRI~\cite{hoopes2022synthstrip}. We show that although the addition of synthetic labels is helpful, simple geometric shapes alone are insufficient to train a model that accurately discriminates between the fetal brain and other fetal and maternal anatomy, resulting in a high number of false-positive voxel predictions. Motivated by state-of-the-art methods that break up the brain-extraction task into a localization step followed by a binary segmentation step~\cite{salehi2018real,ebner2020automated}, we propose a test-time search strategy employing the classical sliding-window approach \cite{cascade15} to reduce the number of largely uncorrelated false positives and to progressively refine the brain-mask prediction within a region of interest.

In contrast to works that apply networks for binary classification of sliding windows, for example, for face detection~\cite{cascade15,slidingwindow2009,fu2022tf}, we adapt the approach for voxel-level segmentation. First, we perform a breadth-fine search (BFS) and systematically explore the full FOV. Second, we narrow the search frame down to a tight region centered on the fetal brain, which facilitates the distinction between brain and surrounding anatomy. Third, a deep focused sliding window (DFS) search results in a cascade of segmentations that we fuse to obtain a brain mask. The method employs multiple models, each trained at a different window size, using synthetic images that span full, partial, and absent fetal brains to capture variability in head sizes and FOV positioning.



We evaluate our approach for fetal brain extraction using two clinical, full-uterus stack-of-slices MRI datasets, one acquired with Half-Fourier Acquisition Single-Shot Turbo Spin Echo (HASTE), the other with Echo Planar Imaging (EPI).
Our framework matches or exceeds the performance of state-of-the-art methods trained on real images.
These results highlight the utility of the sliding window approach in DL and of combining predictions from multiple networks to unlock the potential of synthetic-data training even when only sparse labels are available for image generation. Our code and models are available on GitHub\footnote{\url{https://github.com/dadashkarimi/cascade}}.

\section{Method}

Our framework uses three-dimensional (3D), multi-scale deep neural networks to derive a fetal-brain mask from a full-uterus stack-of-slices scan $I$. We separately train four models---$A$, $B$, $C$, and $D$---to extract the fetal brain from 3D patches of decreasing size, synthesizing training images from label maps of the fetal brain augmented with random geometric shapes~\cite{hoffmann2021synthmorph}. At inference, we pool information from sliding-window predictions of varying size to identify a region of interest (ROI) $R \subset I$ that most likely contains the fetal brain, refine it, and derive the fetal-brain mask $S_\text{final}$.

\begin{figure*}[t]
    \centering
    \includegraphics[width=\textwidth]{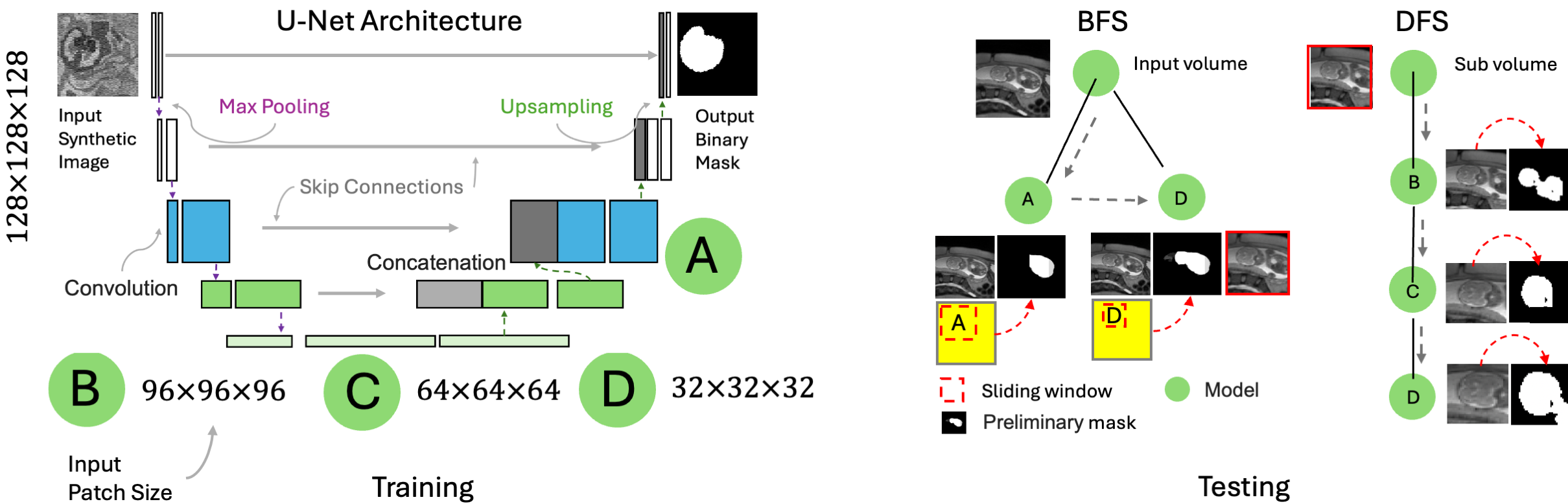}
    \vspace{-0.8cm}
\caption{Proposed BFS/DFS search strategy for fetal-brain extraction. \textbf{BFS}: At test, U-Nets trained on large ($A$) or small cubic patches ($D$) search the full slice stack in a sliding windows fashion to localize the brain. We proceed with DFS within a bounding box (red) fitted to the largest connected component of non-zero output probabilities. \textbf{DFS}: U-Nets trained on progressively smaller cubic patches ($B$--$D$) segment the bounding box patch-wise. We fuse their predictions by majority vote.}
\vspace{-0.3cm}
    \label{fig:pipeline}
\end{figure*}

\subpara{Breadth-Fine Search}

The first inference stage employs Models $A$ and $D$, trained on cubic image volumes of side length $w$ (Table~\ref{tab:label_map_params}). At test, these models process the full-size image $I$ in a sliding-window fashion with step size $s$ (Table~\ref{tab:label_map_params}), generating probability maps $P_i$ ($i \in \{A, D\}$) that indicate the likelihood that voxels contain fetal brain.
We construct a binary mask as the largest connected component of non-zero voxels in any $P_i$, that is, from $P = \{p \in P_i, \forall i : p > 0\}$, fit a bounding box, and extract the corresponding 3D ROI $R$ from the full-size input stack. Next, we perform a deep focused sliding window search to refine $R$.




\subpara{Deep-Focused Sliding Window}
\label{sec:dfs}

The second stage uses Algorithm~\ref{alg:dfs} with $N=3$ models, trained as for BFS.
First, a sliding-window pass with Model $B$ across ROI $R$ leads to probability map $P_B$, which we threshold at $\alpha = 0.2$ to create binary mask $S_B$. Second, we update $R$ to be the bounding box of image $I$ that corresponds to the largest connected component of $S_B$.
Third, we repeat the same steps with Models $C$--$D$, progressively narrowing down $R$ and leading to binary masks $S_C$ and $S_D$.
Finally, we reconstruct each $S_i$ to a full-size map $S_i^\text{full}$, matching the size of input $I$, and obtain the fetal-brain mask $S_\mathrm{final}$ by voxel-wise majority vote across all $S_i$, for $i \in \{B, C, D\}$.
The search strategy manages false positives in that it requires several models to flag a voxel as brain for it to be included in $S_\mathrm{final}$ and by progressively excluding image regions that unlikely contain the fetal brain.



\begin{algorithm}
\caption{DFS strategy for $N$ models}
\label{alg:dfs}
\begin{algorithmic}[1]
\State \textbf{input:} image $I$, search region $R$, models $M_i$, window sizes $w_i$, step sizes $s_i$ ($i \in \{1, 2, ..., N\}$), threshold~$\alpha$
\State \textbf{output:} binary fetal-brain mask $S_{\text{final}}$

\For{$i = 1 \to N$}
    \State Initialize empty volume $P_i$ of size of $R$
    \State Find window locations within $R$ using $w_i$ and $s_i$
    \For{\textbf{each} window $x \subset R$}
        \State $y \gets \text{model}~M_i(I(x))$ \Comment{Predict}
        \State $P_i(x) \gets P_i(x) + y$ \Comment{Accumulate}
    \EndFor
    \State $S_i \gets \{p \in P_i : p \geq \alpha\}$ \Comment{Binarize volume}
    \State Fit bound.\ box $B$ to largest conn.\ component of $S_i$
    \State $R \gets B$ \Comment{Update region}
    \State $S_i^\text{full} \gets \text{recon}(S_i, I)$ \Comment{Zero-pad to size of $I$}
\EndFor

\State $S_{\text{final}} \gets \text{majority\_vote}(S_1^\text{full}, ..., S_N^\text{full})$
\State \textbf{return} $S_{\text{final}}$
\end{algorithmic}
\end{algorithm}


    


\begin{table}[t]
\centering
\small
\caption{Augmentation and synthesis ranges for training. We uniformly sample translation, rotation (Rot.), and scaling parameters with a maximum amplitude, as well as blurring and noise standard deviations (SD) up to the given value. At test, we use step size $s=64$ for Model $A$ and $s=32$ for $B$--$D$.}
\vspace{2mm}
\begin{adjustbox}{max width=\columnwidth}
\begin{tabular}{lccccccc}
\toprule
\multirow{2}{*}{\textbf{Model}} & \textbf{Window} & \textbf{Added} & \textbf{Shift} & \textbf{Rot.} & \multirow{2}{*}{\textbf{Scale}} & \textbf{Blurring} & \textbf{Noise} \\
\textbf{} & \textbf{Size $w$} & \textbf{Shapes $n$} & \textbf{(mm)} & \textbf{($^\circ$)} & & \textbf{SD (mm)} & \textbf{SD} \\
\midrule
A & 128 & 24 & 48 & 180 & 0.6 & 0.6 & 0.40 \\
B &  96 & 24 & 32 & 180 & 0.4 & 0.4 & 0.20 \\
C &  64 & 24 & 12 & 180 & 0.4 & 0.2 & 0.15 \\
D &  32 &  8 &  6 & 180 & 0.3 & 0.1 & 0.15 \\ 
\bottomrule  
\end{tabular}
\end{adjustbox}
\vspace{-0.3cm}
\label{tab:label_map_params}
\end{table}

\subpara{Label Maps and Image Synthesis}

For model training, we randomly select 40 anatomical 3D brain label maps from the FeTA 2022 dataset~\cite{feta2021}, each of which includes 8 labels for fetal-brain structures and background.
Following prior work~\cite{hoffmann2021synthmorph}, we replace the background with random geometric shapes and synthesize highly variable gray-scale images from the resulting label maps.
First, we draw a FeTA label map and augment it spatially by applying a nonlinear transform including translations, rotations, and scaling. Second, we replace the background label with $n$ random shape labels.
Third, we synthesize a gray-scale image by uniformly sampling a different intensity for every label and by adding Gaussian noise.
Finally, we apply a series of randomized image corruptions like Gaussian blur, a smooth spatial intensity bias field, and downsampling.
The label map augmentation and image synthesis use techniques from SynthMorph~\cite{hoffmann2021synthmorph} with the default sampling ranges, except for those shown in Table~\ref{tab:label_map_params}.

We downsample all FeTA label maps to isotropic 1-mm resolution, for efficient training and to approximate the voxel size of clinical HASTE stacks. We crop or zero-pad them to window size $w_i$ for Model $i \in \{A, B, C, D\}$, centering the fetal brain within the window prior to spatial augmentation.

\subpara{Architecture and Training}
Our models implement 3D U-Nets~\cite{ronneberger2015u} with seven levels, each applying two convolutions with LeakyReLU activation (parameter 0.2), max-pooling, and skip connections from encoder to decoder. The encoder uses [16, 16, 64, 64, 64, 64, 64, 64, 64, 64, 64] filters, while the decoder mirrors this structure with [64, 64, 64, 64, 64, 64, 64, 64, 64, 16, 16, 2] filters. The final layer applies a two-filter convolution with softmax activation for probabilistic output of brain and non-brain structures. We optimize model parameters using Adam with learning rate $10^{-5}$, batch size 1, and a soft Dice loss between model predictions and one-hot ground-truth brain labels until the loss plateaus.

\section{Experiments}

We compare our method to state-of-the art fetal-brain extraction baselines~\cite{salehi2018real,ebner2020automated,keraudren2014automated}, using model weights distributed by the respective authors, to gauge the performance attainable by end users without retraining.

\subpara{Evaluation Data}

The experiments use full-uterus stacks of MRI slices with corresponding manual 3D brain masks:
1) A clinical dataset of 70 T2-weighted HASTE stacks from Massachusetts General Hospital (MGH), with $\sim$1 mm\textsuperscript{2} in-plane resolution and 3--4 mm slice thickness. Of these, 35 scans span 13--26 gestational weeks (GW) and 35 span 26--37 GW.
2) A set of 29 T2*-weighted 2D-EPI stacks of 3-mm slices with 3×3 mm\textsuperscript{2} in-plane resolution from Boston Children's Hospital (BCH), balanced between the second and third trimesters~\cite{hoffmann2021rapid}.
We resize images and brain masks to 1-mm isotropic resolution, using linear and nearest-neighbor interpolation, respectively, and symmetrically crop and zero-pad to obtain $192^3$ volumes. This ensures that any anatomy has the same scale relative to the voxel size as at training, while isotropic voxels facilitate 3D convolutions.

\subpara{Baseline Methods}

We test fetal-brain extraction methods that provide a readily usable implementation online and are applicable to unprocessed images, as opposed to super-resolution reconstructions from several HASTE scans.
First, we assess ``Loc-Net''~\cite{ebner2020automated}, a two-step approach performing brain localization followed by segmentation using a 2D P-Net.
Second, we test ``RF/CRF''~\cite{keraudren2014automated}, a method leveraging handcrafted, bundled SIFT features with a Random-Forest classifier and Conditional Random Field (CRF).
Third, we consider a ``2D U-Net'' with autocontext~\cite{salehi2018real}.
For comparison, we also evaluate a sliding-window pass of Model~$A$, as well as a single forward pass of a model trained on full-size synthetic $192^3$ volumes (``192-Net'', no sliding window at test).
While the baseline methods focus on HASTE, the test data are out-of-distribution for all methods, as none---including ours---train on the specific MGH HASTE or any EPI scans.


\begin{figure}[ht!]
    \centering
    \includegraphics[width=\linewidth]{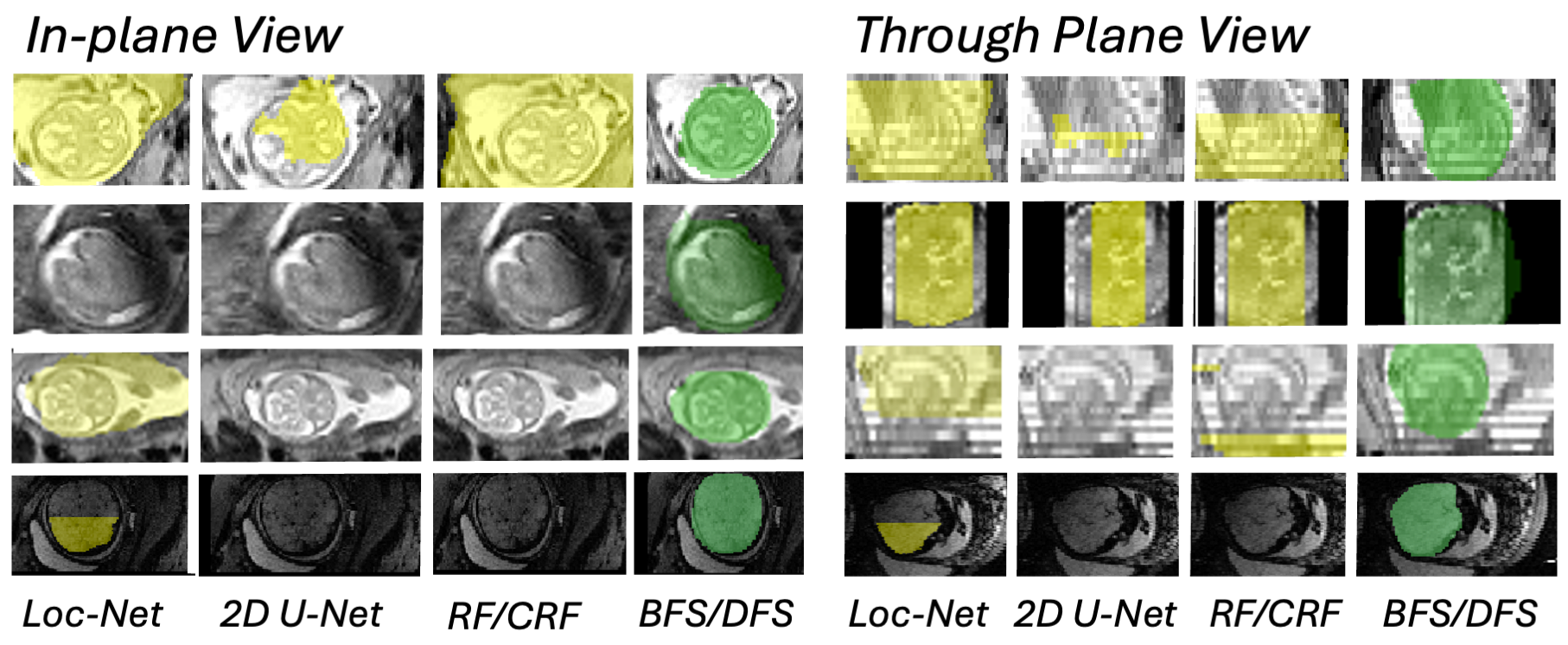}
    \vspace{-0.8cm}
    \caption{
    Representative brain-mask examples for baselines (yellow) and our method (g.) on 4 cases, 1 per row. For each fetus, we show views within and across the imaging plane.}
    \label{fig:examples}
\end{figure}

 \begin{figure*}[t]
    \centering
    \includegraphics[width=\textwidth]{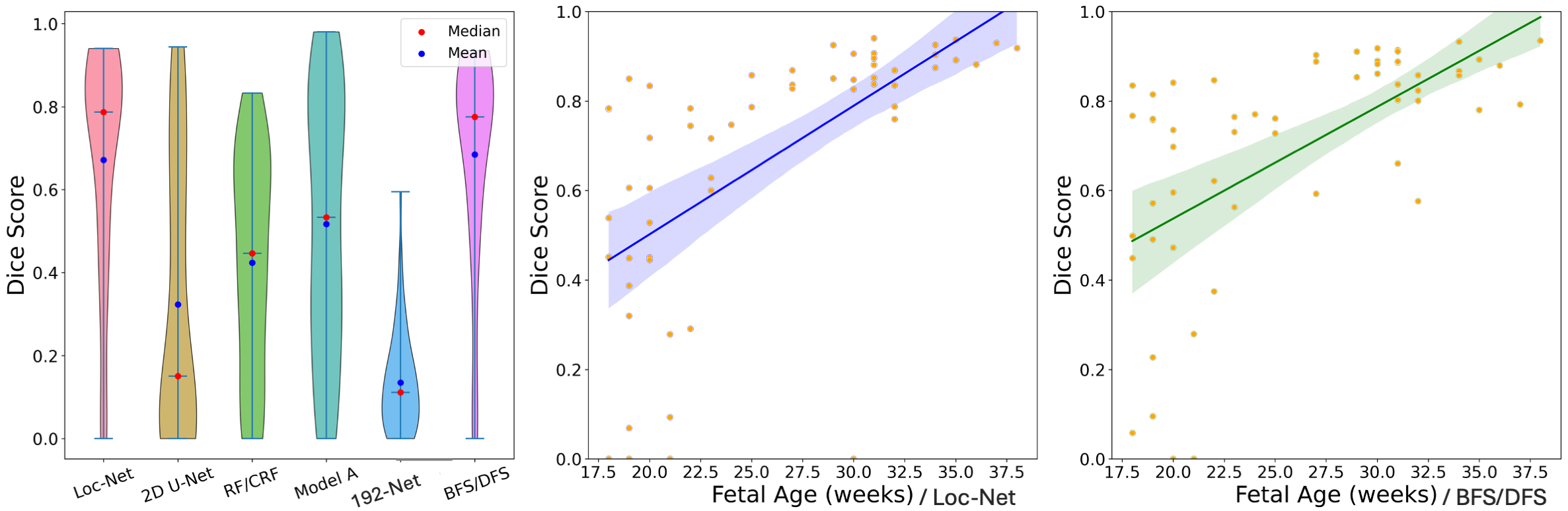}
    \vspace{-8mm}
    \caption{
        Brain-extraction accuracy across the HASTE dataset. 
        \textbf{(left)}~Distribution for all baselines and BFS/DFS (ours).
        Younger fetuses tend to be more challenging, as regression lines of accuracy on age show for the best-performing methods Loc-Net \textbf{(middle)} and BFS/DFS \textbf{(right)}.
        Individual points represent 3D Dice overlap between predicted and ground-truth brain masks.
    }
    \label{fig:results}
    \vspace{-2mm}
\end{figure*}

\newcommand{\starspace}{\hphantom{$^*$}}
\begin{table*}[ht!]
\centering
\small
\caption{Breakdown of brain-masking accuracy in terms of mean 3D Dice overlap by dataset and trimester. The asterisks denote significant differences from Loc-Net at the $p < 0.05$ level, based on paired t-tests.}
\vspace{2mm}
\begin{adjustbox}{max width=\linewidth}
{
\begin{tabular}{lccccccc}
\toprule
&\textbf{Gestation (Weeks)} & \textbf{Loc-Net} & \textbf{2D U-Net} & \textbf{RF/CRF} & \textbf{Model $A$} & \textbf{192-Net} & \textbf{BFS/DFS (Ours)} \\
\midrule
\multirow{3}{*}{\rotatebox{0}{HASTE}} &  18--37    &  0.6714 & 0.3235 & 0.4238 & 0.5170 & 0.1352 & \textbf{0.6881}\starspace \\ 
&  18--25    & 0.5115  & 0.2069  & 0.2395  & 0.3846 & 0.0654 & \textbf{0.5369}$^*$  \\
&  26--37    & {\textbf{0.8428}}  & 0.4484  & 0.6213  & 0.6589 & 0.2100 & {0.8427}\starspace \\
\midrule
\multirow{1}{*}{EPI} &  21--38   & 0.5646  & 0.0066  & 0.0060  & 0.159 & 0.0576 & \textbf{0.5949}$^*$ \\
\bottomrule
\end{tabular}
}
\vspace{-2cm}
\end{adjustbox}
\label{tab:breakdown}
\end{table*}

\subpara{Setup}

We use 5 second-trimester and 5 third-trimester HASTE stacks for model development and validation. After excluding two high-motion scans, we test on 58 held-out HASTE and all available EPI scans, normalizing intensities as expected by each method.
We evaluate brain-extraction accuracy in terms of the Dice coefficient~\cite{dice1945measures} between predicted and manual ground-truth masks.

\subpara{Results}

Figure~\ref{fig:examples} shows several brain-extraction examples. Figure~\ref{fig:results} compares the distribution of brain-masking accuracy across all HASTE stacks, and Table~\ref{tab:breakdown} provides a breakdown by GW as well as results for EPI stacks.

Across all HASTE stacks, the proposed BFS/DFS method achieves the highest Dice score, 0.688, matched by Loc-Net (0.671, $p > 0.05$ for paired t-test). Both methods perform significantly better than RF/CRF (0.424, $p < 0.05$) and 2D U-Net (0.324, $p < 0.05$).
Considering the second-trimester HASTE stacks separately (18--25 GW), BFS/DFS outperforms all baselines tested (0.537, $p < 0.05$). This GW range is generally more challenging, as the smaller size of the fetal brain, limited contrast, and more severe artifacts tend to impinge on accurate masking.
Conversely, improved contrast and less motion tend to render brain masking an easier task in the third trimester (Table~\ref{tab:breakdown}, 26--37 GW), as demonstrated by the regression lines of accuracy on GW for the best-performing methods BFS/DFS and Loc-Net (Figure~\ref{fig:results}).
The sliding-window pass of Model~$A$ and 192-Net considerably lag behind BFS/DFS in each GW cohort, indicating that random-geometric shape labels alone are an insufficient replacement for missing labels of non-brain anatomy. Clearly, pooling predictions from several models trained at different patch sizes using the BFS/DFS strategy substantially enhances accuracy.
For the lower-resolution EPI data which none of the methods---including ours---sampled at training, BFS/DFS achieves the highest Dice score of 0.595, surpassing Loc-Net (0.565).


\section{Discussion and Conclusion}

In this study, we introduce a hybrid search strategy for fetal brain extraction from clinical full-uterus MRI, leveraging networks we train on images synthesized from sparse anatomical labels with added geometric shapes. By merging BFS searches with several models and progressively refining brain masks with DFS, our approach significantly reduces false-positive voxel predictions compared to a single model trained the same way. Our method matches the accuracy of state-of-the-art baselines for HASTE in the third trimester while exceeding their Dice scores by up to 5\% in the second trimester and for EPI across both trimesters. These findings demonstrate the effectiveness of pooling information from the predictions of several models for enhancing brain-extraction accuracy when only sparse labels are available for synthesis-based training.

Our results align with previous research indicating that training on images of arbitrary contrast enables better generalization across diverse real image types than methods reliant on datasets of comparatively low variability~\cite{hoopes2022synthstrip,kelley2024}. By increasing the utility of the synthesis to training sets with sparse labels, we address the need for DL models that can generalize across populations and imaging contrasts. In the future, we will compare our 3D method to BFS/DFS using 2D CNNs, in the presence of slice-by-slice fetal motion. While the baselines tested process each slice individually to avoid incoherent image geometries resulting from between-slice motion, we choose 3D convolutions to share information across slices, which we find to perform robustly for the low to medium-motion cases tested.



\section{Acknowledgments}

The research was supported in part by NIH/NICHD grants R00 HD101553, R01 HD109436, R01 HD102616, and R21 HD106038. 
The project benefited from computational hardware provided by the Massachusetts Life Sciences Center. MH is an advisor to Neuro42, Inc. His interests are reviewed and managed by MGH and Mass General Brigham in accordance with their conflict-of-interest policies.

\section{Compliance with ethical standards}

For the MGH and BCH data acquisition, institutional protocols were followed for image acquisition.
We signed data use agreements for access to FeTA 2022 data.

\bibliographystyle{IEEEbib}
\bibliography{refs}

\end{document}